\documentclass[runningheads]{cl2emult}

\usepackage{makeidx}  
\usepackage{graphicx} 
\usepackage{subeqnar} 
\usepackage{multicol} 
\usepackage{cropmark} 
\usepackage{eso}      
\makeindex            

\newfont{\sfsl}{cmssqi8 scaled 1100}
\newcommand{\gcs}{{\sfsl HIFLUGCS}}
\begin{document}
\title*{Relating Galaxy Cluster\index{Galaxy Clusters} X-Ray
Luminosities
to\,Gravitational\,Masses\,in\,Wide\,Angle\,Surveys}
\toctitle{Relating Galaxy Cluster X-Ray Luminosities
to Gravitational Masses in Wide Angle Surveys}
%
%
\titlerunning{Relating X-Ray Luminosities to Gravitational Masses}
%
\author{Thomas H. Reiprich \and Hans B\"ohringer}
\authorrunning{T.H. Reiprich \& H. B\"ohringer}
%
%
\institute{Max-Planck-Institut f\"ur extraterrestrische Physik, P.O. Box
1312,\protect\newline 85741 Garching, Germany; reiprich@mpe.mpg.de,
hxb@mpe.mpg.de}

\maketitle              

%

\section{Introduction}
The cosmologically most important cluster parameter is its gravitational mass.
In order to constrain cosmological models of structure formation by comparing
simulations or analytic models with, e.g., observational mass functions of
statistical cluster samples,
cluster masses are needed. Precise cluster mass determinations
for large cluster surveys, e.g.~\cite{bs00}, however, require extensive observations and are
currently not feasible. An alternative approach is to use established relations
between more easily observed quantities and cluster mass. 
For optical cluster samples the observed number of cluster galaxies has
sometimes been used to relate to cluster mass. For
X-ray cluster samples a three component approach is commonly used. First, 
measured X-ray luminosities are
converted to X-ray temperatures using an observational relation. 
Then, a slope for the X-ray temperature -- gravitational mass relation is assumed
on theoretical grounds. And finally, the X-ray
temperatures are converted to masses by normalizing this relation with
hydrodynamical simulations.

Here we show that cluster surveys selecting clusters by their X-ray luminosities
effectively select by cluster mass. It is shown that a selection based solely on
Abell galaxy richness is less efficient in terms of mass.
Furthermore we quantify the empirical X-ray
luminosity -- gravitational mass relation and its scatter, offering the
possibility to
directly use this observed relation in wide angle X-ray cluster surveys for the
luminosity -- mass conversion. Throughout $H_0=50\,\rm km/s/Mpc$,
$q_0=0.5$ and $\Lambda=0$ is used.

\section{Results and Discussion}

\begin{figure}
\centering
\includegraphics[width=.95\textwidth]{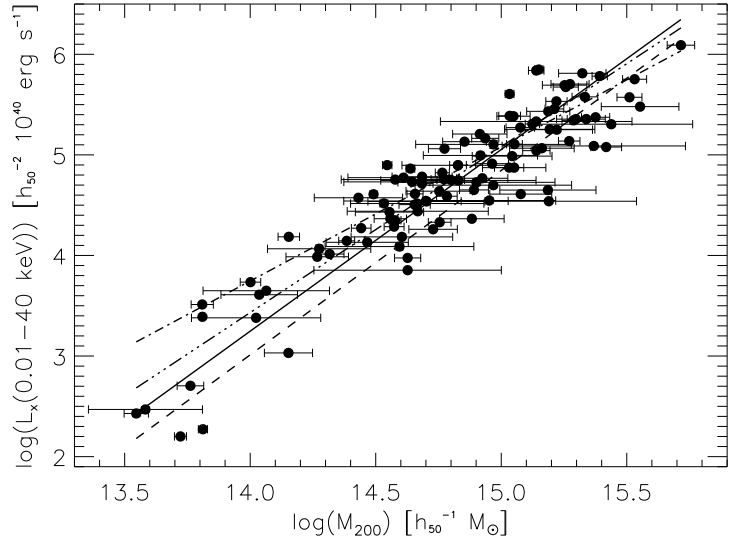}
\caption[]{Bolometric X-ray luminosity versus gravitational mass for 106
clusters.}
\label{lbm}
\end{figure}

In the course of constructing a highly complete X-ray flux-limited sample of the
brightest galaxy clusters in the sky (\gcs\index{HIFLUGCS}, \cite{rb99a})
cluster masses for more than 100 clusters have been determined individually from
X-ray data. In
Fig.~\ref{lbm} the bolometric ($0.01-40\,\rm keV$) X-ray luminosity,
$L_{\rm Bol}$, is plottet
versus the gravitational mass within an overdensity of 200 times the critical
density, $M_{200}$.
The solid line shows the best fit relation for 106 clusters, the
triple-dot-dashed line is the best fit relation determined using the 63
clusters
included in the strictly flux-limited sample (\gcs ), the dot-dashed line
is the self-similar relation (slope = 4/3) normalized by the simulations of
Navarro, Frenk \& White \cite{nfw95}, the dashed line is the
`pre-heated'\index{pre-heating} relation given by Evrard \& Henry
\cite{eh91} (slope = 11/6). The normalization given by Evrard \& Henry,
determined
theoretically, agrees with the normalization found in the simulations of
pre-heated clusters by Navarro, Frenk \& White.
The measured relations lie in between the
relations with and without pre-heating. The physical implications will be
discussed in Reiprich et al.\ (in prep.), here we emphasize that the plot shows
measured and predicted $L_{\rm Bol}-M_{200}$ relation to be in rough
agreement. 
Using a bisector linear regression fit routine in log--log space which takes
into account errors in both 
variables and allows for intrinsic scatter \cite{ab96}, the following
best fit relations are found using 106 clusters:
$L_{\rm Bol} =
9.25\times 10^{-23}\,M_{200}^{1.81\pm 0.08}$
and for the luminosity in the ROSAT energy band
$L_{\rm X}(0.1-2.4\,{\rm keV}) =
1.47\times 10^{-19}\,M_{200}^{1.57\pm 0.08}$
($L_{\rm X}$ in units of $10^{40}\,\rm erg/s$ and
$M_{200}$ in solar masses). Note that different best fit values are obtained
when errors in the variables are neglected and only one variable is treated as
dependent (e.g., for the latter relation then a slope of 1.33, in agreement with
the slope of the preliminary $L_{\rm X}-M_{500}$ relation quantified by
Reiprich \& B\"ohringer \cite{rb99b}, is found instead of 1.57). The bisector
method used here ensures 
that variables are treated symmetrically \cite{ifa90}. The 1-$\sigma$ scatter in
log space for the mass around the $L_{\rm Bol} - M_{200}$ relation equals 0.19,
corresponding to a relative mass error of +55\% and $-$35\% when converting
$L_{\rm Bol}$ to $M_{200}$. This error includes intrinsic and measurement
scatter.

\begin{figure}
\hspace{-2.5cm}
\includegraphics[width=9.25cm]{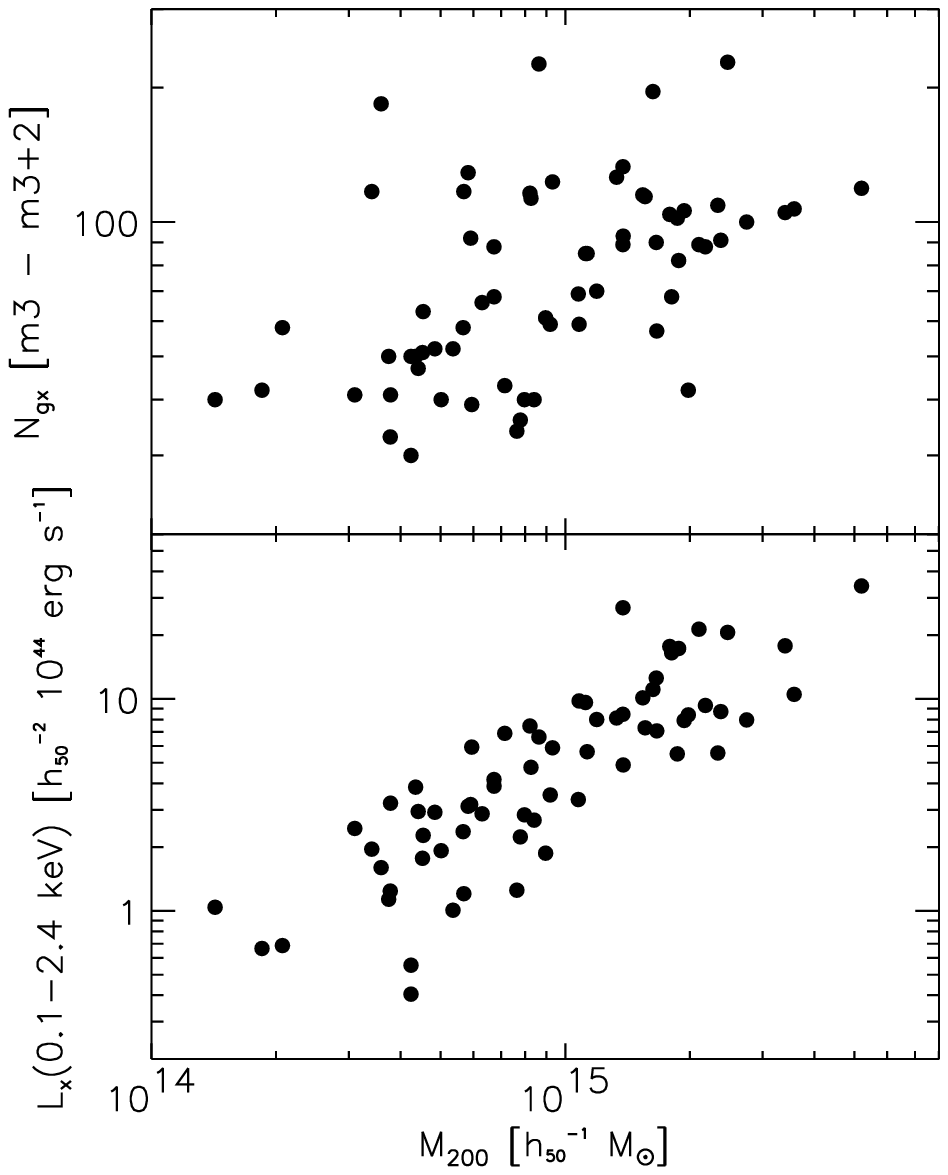}
\hspace{-2.2cm}
\includegraphics[width=9.25cm]{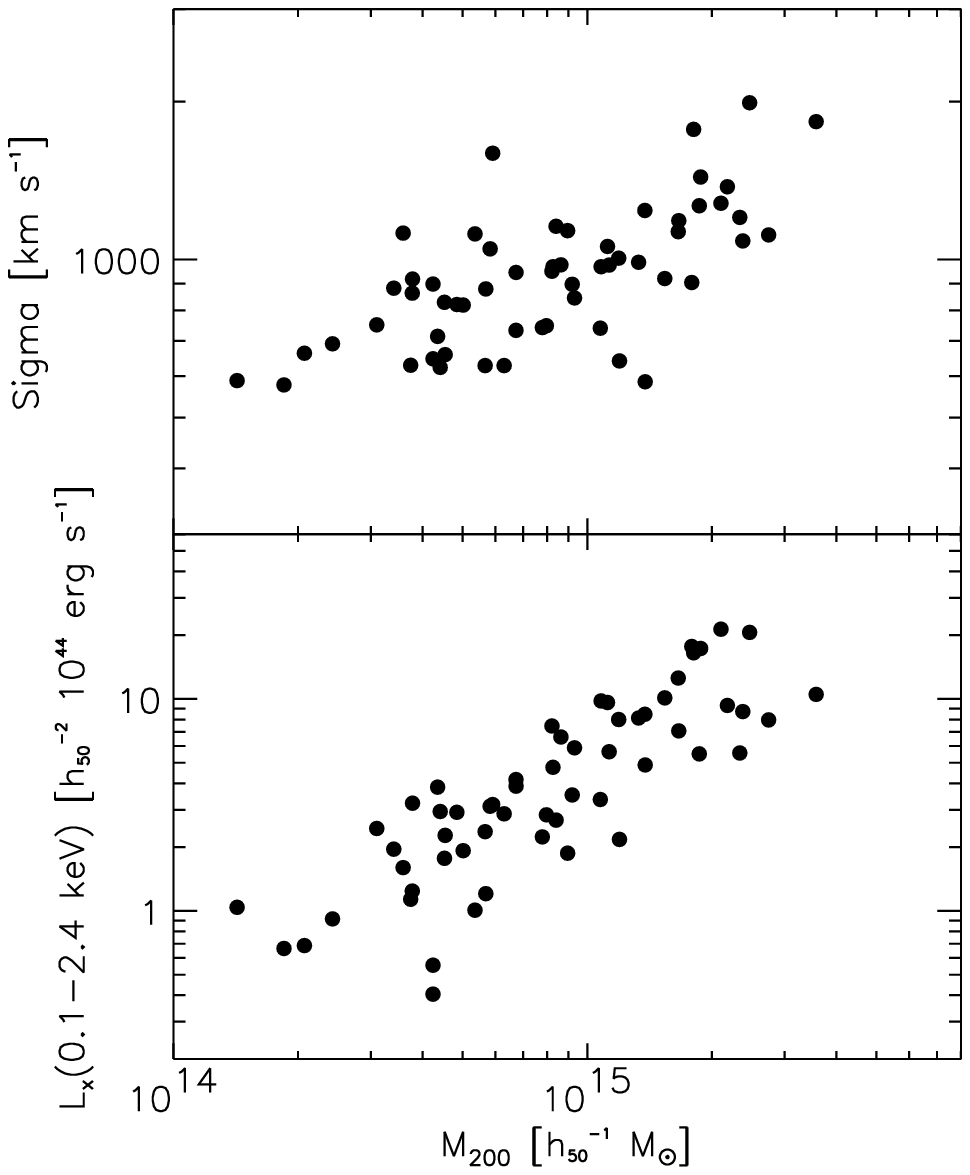}
\caption[]{Number of cluster galaxies for 66 clusters (left) and galaxy
velocity dispersion for 58 clusters (right) versus mass as compared to X-ray
luminosity versus X-ray mass for the same clusters.} 
\label{lr}
\vspace{-0.3cm}
\end{figure}
In Fig.~\ref{lr} the measured number of cluster galaxies, $N_{\rm gx}$, as taken
from Abell, Corwin \& Olowin \cite{aco89} and
the measured radial galaxy velocity 
dispersion, Sigma, as taken from Struble \& Rood \cite{sr99} are compared to $L_{\rm X}$
as gravitational mass tracers.
It is clear from the plot on the left that, assuming $M_{200}$ to be a good
estimate of the true cluster mass, solely taking
the Abell richness to select clusters is not a very effective way to select
clusters by their mass. From the plot on the right one can see that the velocity
dispersion correlates better with mass than the galaxy richness but also it
confirms that careful analyses are required, e.g.~\cite{ggm98}, when determining
the cluster mass from the velocity dispersion.

For further infos and more comprehensive articles please check\\ 
http://www.xray.mpe.mpg.de/$\sim$reiprich/

\vspace{-0.27cm}
\begin{thebibliography}{7}
%
\addcontentsline{toc}{section}{References}

\bibitem{aco89}
Abell, G.~O., {Corwin Jr.}, H.~G., \& Olowin, R.~P. 1989, ApJS, 70, 1

\bibitem{ab96}
{Akritas}, M.~G. \& {Bershady}, M.~A. 1996, ApJ, 470, 706

\bibitem{bs00}
{B\"ohringer}, H. \& {Schuecker}, P., these proceedings

\bibitem{eh91}
{Evrard}, A.~E. \& {Henry}, J.~P. 1991, ApJ, 383, 95

\bibitem{ggm98}
{Girardi}, M., {Giuricin}, G., {Mardirossian}, F., {Mezzetti}, M., \&
  {Boschin}, W. 1998, ApJ, 505, 74

\bibitem{ifa90}
{Isobe}, T., {Feigelson}, E.~D., {Akritas}, M.~G., \& {Babu}, G.~J. 1990, ApJ,
  364, 104

\bibitem{nfw95}
{Navarro}, J.~F., {Frenk}, C.~S., \& {White}, S. D.~M. 1995, MNRAS, 275, 720

\bibitem{rb99b}
{Reiprich}, T.~H. \& {B{\"o}hringer}, H. 1999, Astron. Nachr., 320, 296

\bibitem{rb99a}
{Reiprich}, T.~H. \& {B\"ohringer}, H. 1999, in 19th Texas Symposium on
  Relativistic Astrophysics and Cosmology, ed. J.~{Paul}, T.~{Montmerle}, \&
  E.~{Aubourg}

\bibitem{sr99}
{Struble}, M.~F. \& {Rood}, H.~J. 1999, ApJS, 125, 35

\end{thebibliography}

\clearpage
\addcontentsline{toc}{section}{Index}
\flushbottom
\printindex

\end{document}